\documentclass[10pt,aps,prb,twocolumn,superscriptaddress,nobibnotes,secnumarabic]{revtex4-1}

\usepackage{graphicx}
\usepackage{amsmath}
\usepackage{siunitx}
\usepackage{upgreek}
\usepackage{color}
\usepackage{textcomp}
\usepackage{pdfpages}
\usepackage{pgffor}
\usepackage[tiny,center,compact]{titlesec}
\usepackage[bookmarks=false]{hyperref} 

\hypersetup{
colorlinks=true,
linkcolor=blue,
citecolor=blue,
urlcolor=blue,
hypertexnames=true,
pdfstartview={FitH},				
pdfauthor={Achint Jain}
}

\titlespacing{\section}{0pt}{1.5ex}{1.2ex}					
\makeatletter
\AtBeginDocument{\let\LS@rot\@undefined}						
\makeatother

\begin{document}

\def \mos2{MoS$_{2}$}
\def \sio2{SiO$_{2}$}
\def \uvo3{UV-O$_{3}$}
\def \e12g{\emph{E}{$^{\raisebox{1pt}{$\scriptstyle 1$}}_{\raisebox{-1pt}{$\scriptstyle\text{2g}$}}$}}
\def \a1g{\emph{A}{$^{\raisebox{1pt}{$\scriptstyle$}}_{\raisebox{-1pt}{$\scriptstyle\text{1g}$}}$}}

\newcommand{\rpm}{\raisebox{.1ex}{$\scriptstyle\pm$}}			
\newcommand{\til}{\raisebox{.2ex}{$\scriptstyle\sim$}}
\newcommand{\gt}{\raisebox{.3ex}{$\scriptstyle$\textgreater}}
\newcommand{\doglobally}[1]{{\globaldefs=1#1}}	
\renewcommand{\thesection}{\arabic{section}}		
\renewcommand{\figurename}{\textbf{Figure}}
\renewcommand{\thefigure}{\textbf{\arabic{figure}}}

\title{Minimizing Residues and Strain in 2D Materials Transferred from PDMS} 

\author{Achint Jain}
\affiliation{Photonics Laboratory, ETH Z\"urich, 8093 Z\"urich, Switzerland}

\author{Palash Bharadwaj}
\affiliation{Department of Electrical and Computer Engineering, Rice University, Houston, TX 77005, USA}

\author{Sebastian Heeg}
\author{Markus Parzefall}
\affiliation{Photonics Laboratory, ETH Z\"urich, 8093 Z\"urich, Switzerland}

\author{Takashi Taniguchi}
\author{Kenji Watanabe}
\affiliation {National Institute for Material Science, 1-1 Namiki, Tsukuba 305-0044, Japan}

\author{Lukas Novotny}
\email{lnovotny@ethz.ch}
\affiliation{Photonics Laboratory, ETH Z\"urich, 8093 Z\"urich, Switzerland}

\date{\today}


\begin{abstract}

Integrating layered two-dimensional (2D) materials into 3D heterostructures offers opportunities for novel material functionalities and applications in electronics and photonics. In order to build the highest quality heterostructures, it is crucial to preserve the cleanliness and morphology of 2D material surfaces that come in contact with polymers such as PDMS during transfer. Here we report that substantial residues and up to \til 0.22 \% compressive strain can be present in monolayer \mos2 flakes transferred using PDMS. We show that a UV-ozone pre-cleaning of the PDMS surface before exfoliation significantly reduces organic residues on transferred \mos2 flakes. An additional \SI{200}{\degreeCelsius} vacuum anneal after transfer efficiently removes interfacial bubbles and wrinkles as well as accumulated strain, thereby restoring the surface morphology of transferred flakes to their native state. Our recipe is important for building clean heterostructures of 2D materials and increasing the reproducibility and reliability of devices based on them.

\end{abstract}

\maketitle
\doglobally\small							
\setlength\parskip{-1.5pt}		

\section{Introduction}
2D materials like graphene, hexagonal boron nitride (hBN), transition metal dichalcogenides (TMDCs), etc.\ have gained immense attention in recent years due to the wealth of novel fundamental properties and fascinating physical phenomena exhibited by them \cite{Xu13, Bhimanapati15}. A unique possibility existing with these materials is that of assembling them into 3D heterostructures to create artificial materials which do not exist naturally and thus give rise to new functionalities which seem promising for many future applications \cite{Geim13, Novoselov16}. These heterostructures can be built by a variety of methods \cite{Lim14} like chemical vapour deposition (CVD) \cite{Gong14}, epitaxial growth \cite{Koma85, Lin15}, inkjet printing \cite{McManus17} or more generally by mechanical stacking of individual layers \cite{Dean10, Wang13}. Among the many available techniques for stacking 2D materials \cite{Wang13, Zomer14, Kretinin14, Parzefall15, Pizzocchero16, Frisenda18}, one approach that has become common recently is to mechanically exfoliate bulk 2D crystals onto a stamp made of a viscoelastic material, such as poly-dimethylsiloxane (PDMS), bring them in contact with a desired substrate and then slowly detach the stamp to leave the 2D flakes behind \cite{Meitl06, Gomez14}. Although this procedure is very versatile, deterministic and fairly simple to perform, not much attention has been paid to the surface cleanliness of the flakes transferred this way. \\[-1.5ex]

PDMS is a widely used polymer for contact printing \cite{Kumar93}, microfluidics \cite{Duffy98} as well as stretchable electronics \cite{Rogers10}, and its chemistry has been studied extensively \cite{Xia98}. PDMS is composed of a network of cross-linked dimethylsiloxane oligomers which do not get fully cross-linked even after extensive curing \cite{Regehr09} and depending on the curing time and temperature, up to 5\% of oligomers can remain uncrosslinked within the PDMS bulk \cite{Lee03, Regehr09}. It is well-known that these uncrosslinked species are even present on the surface of PDMS stamps and get transferred to the target substrate during contact printing \cite{Glasmastar03, Briseno06}, thereafter acting as a surface contamination layer. PDMS oligomer residues have been characterized previously by many techniques such as nonlinear spectroscopy \cite{Boehm99}, X-ray photoelectron spectroscopy (XPS) \cite{Glasmastar03, Yunus07}, atomic force microscopy (AFM) and ToF-SIMS \cite{Yunus07}, confirming that PDMS can indeed degrade the surface cleanliness of transferred materials. Residues were also found to occur on 2D materials, \textit{e.g.}\ graphene \cite{Allen09} and TMDCs \cite{Liu14a}, transferred using PDMS. Owing to their two-dimensional structure, the properties of 2D materials are quite sensitive to surface contaminants and residues trapped at the interfaces within 3D heterostructures \cite{Tongay14}. Hence, it is essential to stack these materials with minimum residues.\\[-1.5ex]

In this work, the cleanliness of 2D materials transferred using PDMS was investigated with the help of AFM and photoluminescence (PL) measurements. Using monolayer (1L) \mos2 as a test layer, we show evidence for substantial residues being present on \mos2 transferred onto hBN from PDMS. For transferring \mos2 in a cleaner way, we developed an ultraviolet-ozone (\uvo3) treatment recipe to clean the PDMS surface before exfoliating \mos2 on it and found a significant reduction in residues compared to transfer from untreated PDMS. Using PL and Raman spectroscopy, we further reveal that a small compressive strain can be present in \mos2 after transfer from PDMS. We found that subsequent vacuum annealing leads to an almost pristine \mos2 surface on hBN, mostly free from interfacial bubbles, wrinkles and strain. Although here we chose 1L-\mos2 for demonstration since its sensitive PL and Raman signals allow for systematic optical characterization, our cleaning recipe is general and can be used with other 2D materials as well. \\[-1.5ex]

\section{Results and discussion}

\setlength\belowcaptionskip{-8pt}
\begin{figure*}[t]
\centering
	\includegraphics[width=0.95\textwidth]{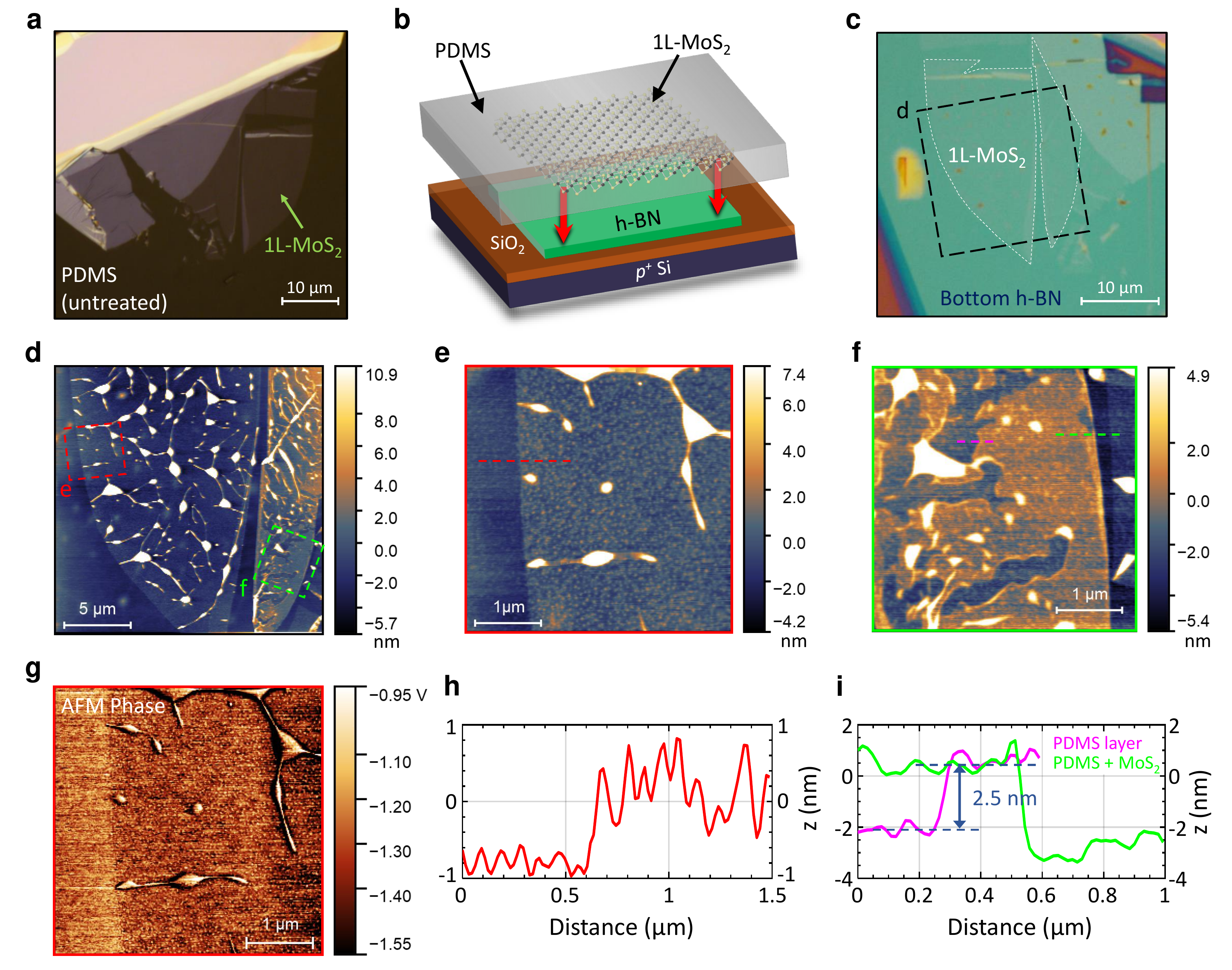}
	\caption{\label{fig1}\textbf{Characterization of PDMS residues. (a)} Optical microscope image of a 1L-\mos2 flake exfoliated on PDMS. \textbf{(b)} Schematic illustration of an inverted PDMS stamp with \mos2 being transferred onto hBN. \textbf{(c)} The same \mos2 flake as in \textbf{a} after transfer to hBN. \textbf{(d)} Large area AFM topography map of the region outlined in \textbf{c} displaying an ``apparently clean'' surface together with some bubbles and wrinkles. \textbf{(e)} Higher resolution AFM map of the red-outlined region in \textbf{d} exhibiting a dense distribution of residues over the entire \mos2 flake. \textbf{(f)} Higher resolution AFM map of the green-outlined region in \textbf{d} with a thick layer of residues covering a majority of the area. \textbf{(g)} AFM phase map recorded together with the topography in \textbf{e} depicting a poor phase contrast between \mos2 and hBN due to PDMS residues on both surfaces. \textbf{(h)} Height profile across the \mos2 edge along the dashed line marked in \textbf{e}. An accurate estimation of the thickness of monolayer \mos2 is hindered by topography variations due to surface residues. \textbf{(i)} Height profile across the PDMS residue layer and PDMS \texttt{+} \mos2 along the dashed lines marked in \textbf{f}. The thickness of the residue layer is \protect\til\SI{2.5}{\nano\meter}.}
\end{figure*}

Naturally occurring \mos2 crystals purchased from SPI Supplies were exfoliated on commercial PDMS films (Gel-Film\textsuperscript{\textregistered} PF-40-X4 sold by Gel-Pak) using a blue tape (BT-150E-KL). Figure 1a is an optical microscope image of a monolayer \mos2 flake on PDMS. Bulk hBN crystals were separately exfoliated on O$_{2}$ plasma cleaned \textit{p$^+$}Si/\sio2 (\SI{285}{\nano\meter}) substrates. The PDMS stamp with \mos2 was placed on a transparent quartz plate and aligned on top of a suitable hBN flake on \sio2 using a mask aligner as depicted schematically in Fig.\ 1b. Upon slowly bringing \mos2 in contact with hBN at room temperature, the entire stack was heated to $\til$\SI{65}{\degreeCelsius} for two minutes using a Peltier module kept underneath the Si/\sio2 substrate. After allowing the stack to cool down, the PDMS stamp was slowly detached as described in ref.\ [\onlinecite{Gomez14}], leaving behind the \mos2 flake on hBN (Fig.\ 1c). \\[-1.5ex]

To characterize the surface of the flakes after transfer, we performed topography mapping using an AFM operating in tapping mode. Figure 1d shows an AFM topography map of the region outlined in black in Fig.\ 1c. The \mos2 flake appears mostly ``clean'' on this large scale apart from the usual wrinkles and bubbles which are frequently seen in PDMS transferred flakes \cite{Gomez14}. However, if we examine the region outlined in red more closely in the higher resolution map in Fig.\ 1e, a dense network of residue islands is clearly visible on the entire \mos2 surface, making it difficult to even obtain an accurate estimate of the 1L-\mos2 thickness from a horizontal cross-section (Fig.\ 1h). Note that these {\rpm}\SI{0.6}{\nano\meter} (rms) variations in topography cannot arise from substrate roughness alone as the hBN layer below is atomically smooth. Moreover, on the narrower 1L-\mos2 flake on the right in Fig.\ 1d, a thick residue layer can be clearly identified which is unmistakably distinct from the \mos2 flake itself. Better visible in the higher resolution AFM map in Fig.\ 1f, this additional layer left-over by PDMS can be as thick as {\til}\SI{2.5}{\nano\meter} in some areas (magenta profile in Fig.\ 1i). Together with the topography map in Fig.\ 1e, we also recorded the corresponding AFM phase map as shown in Fig.\ 1g and observed a poor phase contrast between the \mos2 and hBN surfaces which also indicates the presence of PDMS residues everywhere. \\[-1.5ex]

\setlength\abovecaptionskip{2pt}

\begin{figure*}[t]
\centering
	\includegraphics[width=0.9\textwidth]{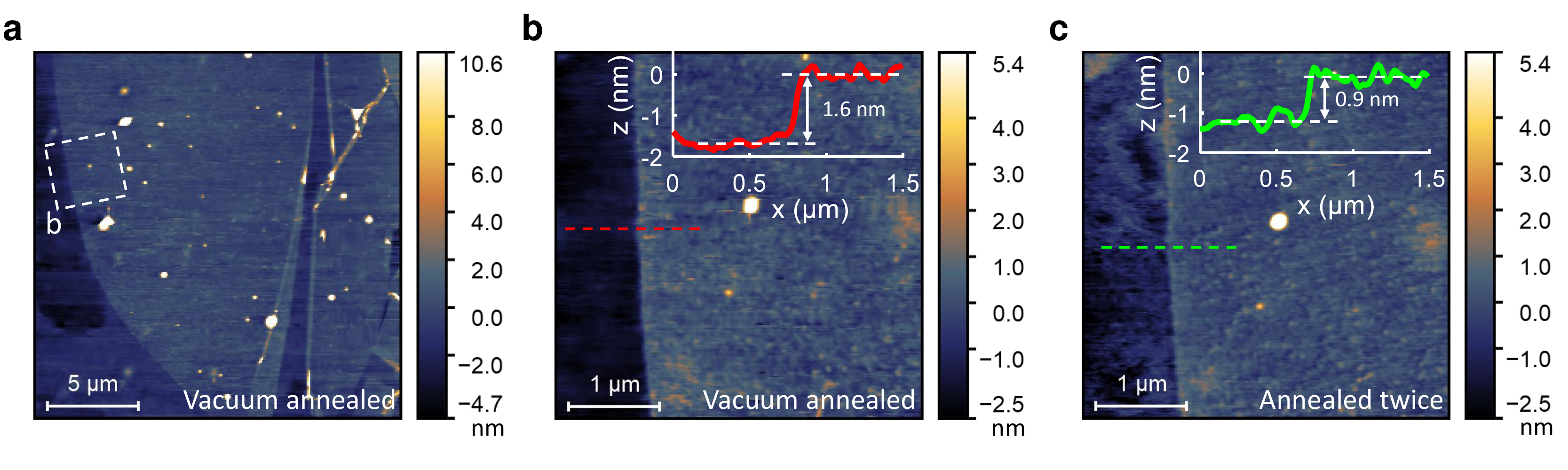}
	\caption{\label{figS1}\small{\textbf{Limited effect of annealing on PDMS residues. (a)} AFM topography map of the dirty 1L-\mos2 flake shown in Fig.\ 1d after \SI{3}{\hour} annealing in vacuum. \textbf{(b)} Higher resolution AFM map of the white-outlined region in \textbf{a}. Even though the surface appears mostly free from bubbles and more homogeneous than before (cf. Fig.\ 1e), a thin layer of PDMS residues is still present over the entire \mos2 flake. Inset: Height profile along the red dashed line revealing a total thickness of \SI{1.6}{\nano\meter} for the monolayer flake. \textbf{(d)} AFM map of the same region after annealing again for \SI{3}{\hour}. Although the total amount of residues does decrease after prolonged annealing, the surface looks far from pristine. Inset: Height profile along the green dashed line.}}
\end{figure*}

We tested several flakes and similar residues were found on all of them (see supplementary Fig.\ S1 for residues present on another flake). The amount of residues transferred was also affected by the temperature and pressure applied during the transfer. In case of transfers done without applying any heat, residues were visible even in an optical microscope due a change in the color of the \sio2 (\SI{285}{\nano\meter}) layer which arises from interference and is sensitive to a few nanometers thick transparent organic layer on top (see supplementary Fig.~S2). These results unambiguously demonstrate, in agreement with previous reports \cite{Briseno06, Boehm99, Glasmastar03, Yunus07, Allen09, Liu14a}, that PDMS can indeed leave a significant amount of residues behind and an efficient method for eliminating them is genuinely needed. \\[-1.5ex]

Although a high temperature (\SI{400}{\degreeCelsius}) vacuum anneal after transfer has been reported to reduce PDMS residues on graphene \cite{Allen09}, such high temperatures are known to introduce additional defects in TMDCs \cite{Tongay13a, Tongay13b} which are more fragile compared to graphene. Moreover, we found that annealing by itself is not sufficient to fully restore the transferred flakes to their pristine state. The \mos2 flake shown in Fig.\ 1c was annealed at \SI{200}{\degreeCelsius} in vacuum for \SI{3}{\hour}. In the AFM maps in Figs.\ 2a, b  it can be noticed that although bubbles and wrinkles are mostly gone, the \mos2 surface is still smeared with residues. This is also evident from the AFM cross-section in Fig.\ 2b revealing the total thickness of the flake to be \SI{1.6}{\nano\meter}, which is significantly higher than the expected value of \SI{0.7}{\nano\meter} for monolayer \mos2. An additional vacuum anneal using the same parameters did somewhat reduce the thickness to \til\SI{0.9}{\nano\meter} as depicted in Fig.\ 2c. However, it can be easily seen that the surface topography is quite inhomogeneous and does not approach the cleanliness level of a freshly exfoliated \mos2 flake.  Alternatively, dissolving PDMS residues in organic solvents such as dichloromethane and toluene \cite{Lee03} is also not very effective as the solvent molecules themselves tend to get adsorbed at exposed 2D material surfaces/edges and can even chemically dope TMDCs \cite{Yang14}. This brings us to the question of how to eliminate PDMS residues in a way that does not compromise the 2D material being transferred in any way. In this regard, it appears more reasonable to clean the PDMS surface itself before it comes in contact with the 2D material. \\[-1.5ex]

To achieve this, we developed a \uvo3 treatment recipe to modify the PDMS surface prior to \mos2 exfoliation. This recipe was found to significantly reduce transfer residues without having any negative effect on the \mos2. \uvo3 cleaning of PDMS has been proposed in the past \cite{Glasmastar03} and the mechanism behind it can be understood as follows. Oxygen free radicals and ozone (O$_3$) created from atmospheric oxygen in the presence of UV-radiation, break down organic species on the PDMS surface into CO$_2$, H$_2$O and simpler volatile organic products. At the same time, the silicon present in poly-dimethyl\underline{sil}oxane gets oxidized and forms a thin (20-\SI{30}{\nano\meter}) layer of silicon oxide (SiO$_{\text{x}}$) on PDMS \cite{Ouyang00}. This SiO$_{\text{x}}$ surface layer besides having a low carbon content, also acts as a diffusion barrier for oligomers still present within the PDMS bulk. \\[-1.5ex]

To optimize the treatment time, we exposed several PDMS stamps to \uvo3 in a Bioforce Nanosciences UV Ozone ProCleaner for varying times, and found a duration of 30-\SI{40}{\minute} to be the optimum. Shorter times did not fully clean the PDMS and longer exposure led to a poorer coverage of flakes on the PDMS upon exfoliation. We also observed that if the exfoliation was done immediately after \uvo3 treatment, bonding often occurred between the PDMS and the blue tape used for exfoliation or in rare cases even between the PDMS and O$_{2}$ plasma treated \sio2 surface during the transfer process. However, this bonding could be avoided by leaving the PDMS exposed to ambient air for a few hours to let the surface undergo a partial hydrophobic recovery \cite{Bodas07, Morent07}. This led us to the following optimized process flow, the results of which are presented below. \\[-1.5ex]

The PDMS stamp was exposed to \uvo3 for \SI{30}{\minute} and then after a wait interval of \SI{2}{\hour} in air, \mos2 was exfoliated on PDMS as usual. Upon optical identification of suitable flakes, transfer was carried out the same way as described earlier. Figure 3a shows an optical microscope image of a large \mos2 flake exfoliated on \SI{30}{\minute} \uvo3 treated PDMS which was subsequently transferred to hBN (Fig.\ 3b). In the AFM topography map of the 1L region shown in Fig.\ 3c, one can notice the absence of dense islands or thick layers of PDMS residues unlike in Figs.\ 1e or 1f. Although small amounts of residues can still be detected, this \mos2 flake is significantly cleaner than the one in Fig.\ 1 which was transferred from untreated PDMS (also see supplementary Fig.\ S3 for better resolved AFM images of another clean \mos2 flake). \\[-1.5ex]

\begin{figure*}[t]
\centering
	\includegraphics[width=0.9\textwidth]{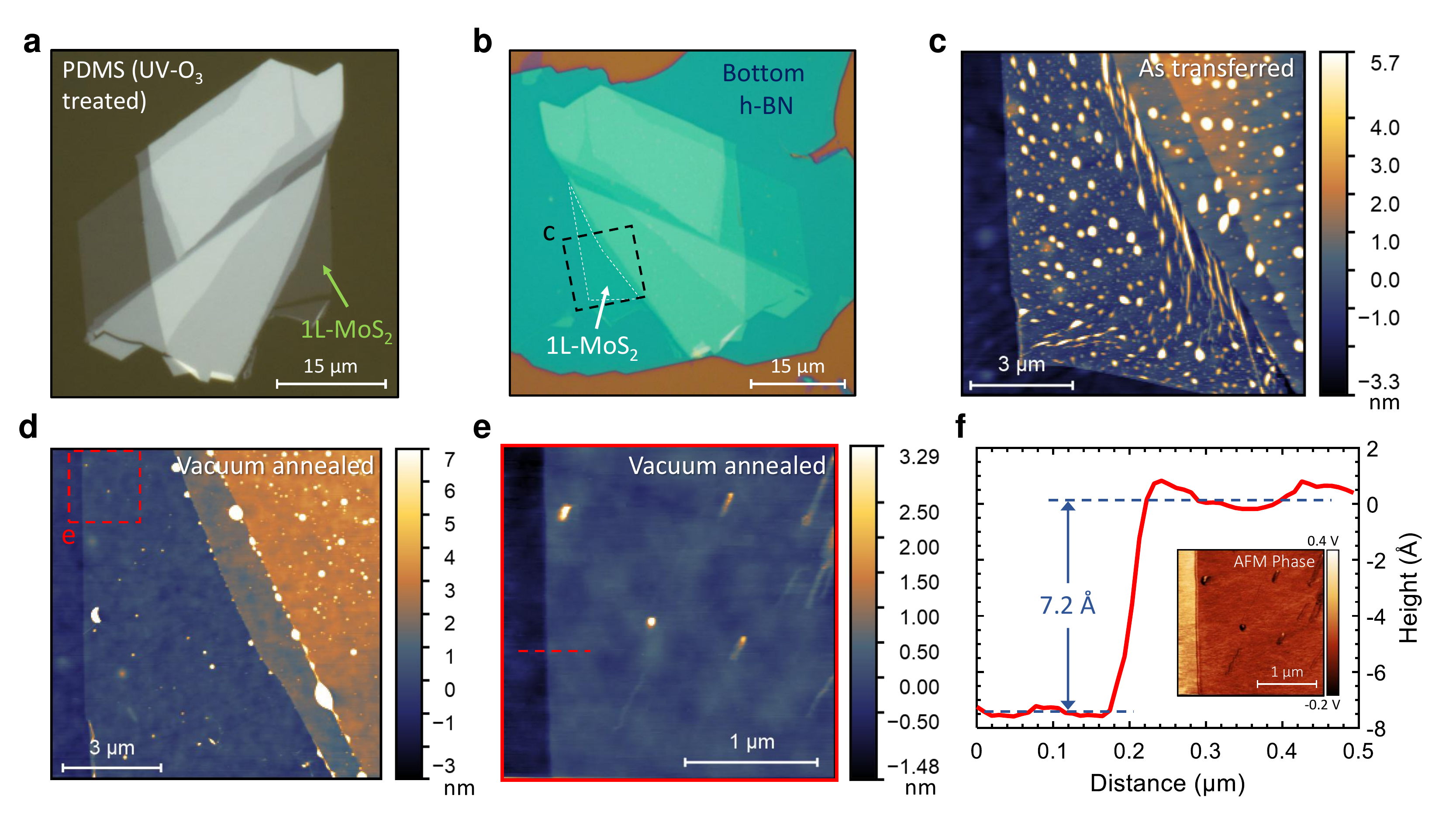}
	\caption{\label{fig2}\textbf{Clean transfer from UV-ozone cleaned PDMS. (a)} Optical microscope image of 1L-\mos2 exfoliated on \SI{30}{\minute} \uvo3 cleaned PDMS. \textbf{(b)} The same \mos2 flake after transfer to hBN with the monolayer segment demarcated. \textbf{(c)} AFM topography map of the region outlined in \textbf{b}. Areas densely covered with residues like in Figs.\ 1e or f could no longer be detected although a considerable amount of bubbles do occur which could also be efficiently removed. \textbf{(d)} AFM map of the same region showing a drastic reduction in bubbles and wrinkles from the entire flake after \SI{200}{\degreeCelsius} vacuum annealing. \textbf{(e)} Higher resolution AFM map of the red-outlined region in \textbf{d} displaying a very clean and smooth \mos2 surface. The faint streaks correspond to migration paths of residual bubbles which couldn't reach the \mos2 edge and get released. \textbf{(f)} Height profile along the red-dashed line in \textbf{e} exhibiting a clear monolayer \mos2 step of \SI{7.2}{\angstrom} unlike in Figs.\ 1h or 2b-c. Inset: AFM phase map recorded together with the topography in \textbf{e} revealing a clear phase contrast between \mos2 and hBN which further indicates the absence of residues (cf. Fig.\ 1g).}
\end{figure*}

The large number of bright spots in Fig.\ 3c are wrinkles and bubbles filled mainly with air molecules and organic adsorbates that become trapped between \mos2 and hBN during transfer and get squeezed into small pockets via a self-cleaning effect \cite{Kretinin14, Haigh12}. The amount of trapped bubbles can be reduced by performing the exfoliation and transfer inside an Ar filled glove box to exclude molecular adsorbates \cite{Cao15} or even fully eliminated by stacking in vacuum as shown recently \cite{Kang17}. Alternatively, we found that bubbles and wrinkles can be very efficiently removed by vacuum annealing at \SI{200}{\degreeCelsius}, similar to another recent report \cite{Wierzbowski17}. At this temperature and under low pressure, the trapped species inside the bubbles become mobile and coalesce into bigger bubbles to minimize the total surface energy \cite{Frisenda18}. They also tend to migrate towards the edge from where they eventually escape the interface, thereby lowering the overall density of bubbles. Figure 3d is an AFM map of the same region as Fig.\ 3c after \SI{3}{\hour} vacuum annealing at \SI{200}{\degreeCelsius} exhibiting a complete removal of wrinkles and a remarkable reduction in bubbles. The higher resolution AFM map in Fig.\ 3e of the red outlined region features a nearly pristine surface with a clean step of \SI{7.2}{\angstrom} (Fig.\ 3f) and a surface roughness of \rpm\SI{1.4}{\angstrom} (rms) over the entire flake. This is in striking contrast to the \mos2 flake transferred from untreated PDMS where the surface quality remained compromised by PDMS residues even after prolonged annealing as shown in Figs.\ 2b-c. The corresponding AFM phase map in Fig.\ 3f (inset) shows a clear phase contrast between \mos2 and hBN (unlike Fig.\ 1g) providing further evidence for the decrease in residues by \uvo3 treatment of PDMS. Detailed AFM analysis of two additional 1L-\mos2 flakes transferred from \uvo3 treated PDMS can be found in supplementary Figs.\ S3-S4 showing similarly clean surfaces. \\[-1.5ex]

Thus our recipe provides a new method to obtain very clean 2D material flakes on hBN. We would like to stress that a combination of \uvo3 pre-cleaning followed by vacuum annealing is crucial, and vacuum annealing by itself does not result in a pristine surface as evident from Fig.\ 2. Note that here we transferred \mos2 onto atomically smooth hBN flakes to decouple the \sio2 substrate roughness from the surface topography of \mos2 which allowed us to better resolve surface residues using AFM. Although the surface cleanliness doesn't depend on the substrate used, we have observed that the removal of bubbles and wrinkles during vacuum annealing is not very effective for \mos2 transferred directly on \sio2 (see discussion in supplementary section S4). \\[-1.5ex]

\begin{figure*}[t]
\centering
	\includegraphics[width=0.9\textwidth]{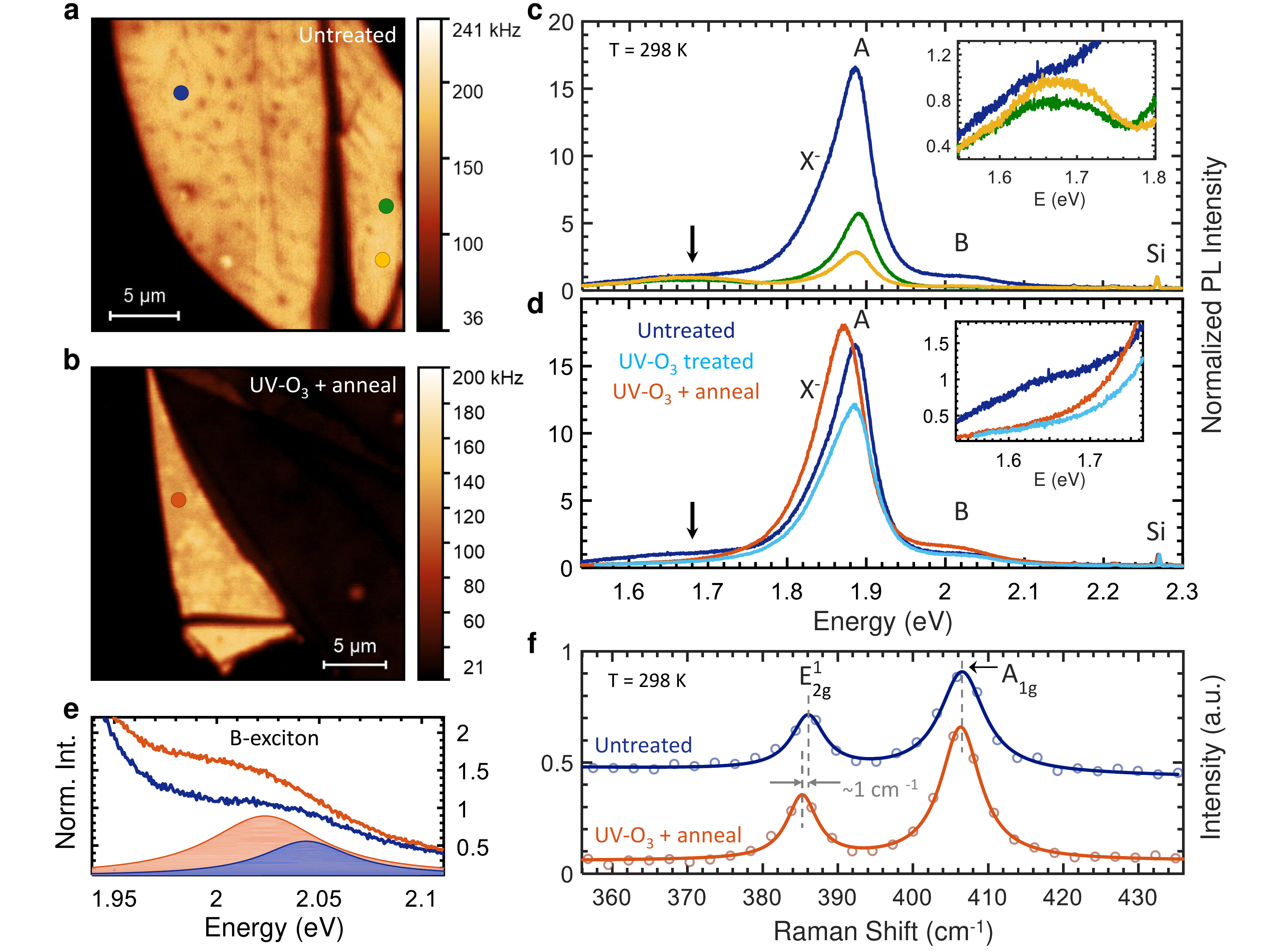}
	\caption{\label{fig3}\textbf{Optical characterization. (a,\ b)} Spatial maps of the spectrally integrated PL counts from the 1L-\mos2 flakes in Figs.\ 1d,\ 3d displaying similar emission intensities. The dark spots and lines in \textbf{a} are interfacial bubbles and wrinkles respectively (cf.\ Fig.\ 1d) where the photon counts were somewhat decreased. \textbf{(c)} PL spectra of \mos2 transferred from untreated PDMS recorded at points marked by matching colored dots in \textbf{a}. Besides the indicated excitonic peaks, all spectra also show a low-energy feature (marked by arrow) around \SI{1.68}{\electronvolt}. \textbf{(d)} PL spectra of \mos2 transferred from \uvo3 treated PDMS showing the absence of any low-energy feature in comparison with \mos2 from untreated PDMS. This indicates that the new feature arises only in the presence of PDMS residues on \mos2. Apart from this, the PL spectra in light and dark blue show blue-shifted A, B excitonic peaks which reveals the presence of compressive strain in as-transferred \mos2 (irrespective of PDMS pre-treatment). This strain got released from the flake in \textbf{b} upon annealing. The PL spectra in light blue was recorded from the flake in Fig.\ S3a before annealing while the annealed \mos2 spectra (orange) was taken from the flake in \textbf{b}. Insets: magnified plots of the spectra around \SI{1.68}{\electronvolt}. \textbf{(e)} A magnified view of the B-exciton emission in \textbf{d}. The filled area plots are Lorentzian fits to the B-exciton peaks and display a clear shift in the center positions. \textbf{(f)} Raman spectra of the two flakes (vertically offset for clarity). The empty circles are measured data points while the smooth lines are Lorentzian fits to estimate the center position of each peak. The \protect\e12g peak (in blue) is up-shifted by \protect\til\SI{1}{\per\centi\meter} which also indicates a build-up of compressive strain in as-transferred \mos2, in agreement with the PL spectra. All spectra have been normalized to the silicon Raman peak for a better comparison.}
\end{figure*}

\section{Optical Characterization}

To further characterize the effect of PDMS transfer on the optical properties of \mos2, we performed PL and Raman spectroscopy. Experimental details can be found in the \emph{Methods} section. Spatial maps of the spectrally integrated PL counts were recorded from the \mos2 flake in Fig.\ 1d transferred using untreated PDMS (labelled as `untreated' in Fig.\ 4) as well as from the clean \mos2 flake transferred using \uvo3 treated PDMS and subsequently vacuum annealed (Fig.\ 3d). PL maps of the two flakes depicted in Figs.\ 4a,\ b show no apparent differences and the count rates were comparable for similar excitation powers. This indicates that PDMS residues (and/or vacuum annealing) do not seem to affect the overall PL quantum yield which could be one reason why residues have been overlooked in the past. The PL spectra of the two flakes, however, did show some interesting differences. \\[-1.5ex]

Besides the well-known A, B-exciton and trion (X$^-$) peaks, a new broad feature around \SI{1.68}{\electronvolt} can be noticed in Fig.~4c in the PL spectra of untreated \mos2. This low energy peak can be better visualized at the locations marked by green and yellow dots in Fig.\ 4a where thick PDMS layers were present on \mos2 (as shown previously in the AFM map in Fig.\ 1f). At these spots, the main excitonic peaks were weaker which made it easier to resolve the new peak (see Fig.\ 4c inset). Surprisingly, this peak could not be detected in \mos2 flakes transferred from \uvo3 treated PDMS (light blue and orange curves in Fig.\ 4d inset) as well as in the fluorescence spectra of an untreated PDMS stamp without \mos2 (data not shown here). It appeared in the PL spectra of \mos2 only when PDMS residues were present on \mos2 but its true origin is unclear at present. Such a broad low energy emission could possibly be attributed to impurity bound excitons at room temperature \cite{Tongay13b, Chow15} whose creation, however, is still not well understood in literature and a more detailed investigation is needed to elucidate the exact mechanism that gives rise to this additional peak. \\[-1.5ex]

Apart from this, it can be clearly observed in Fig.\ 4d that the A-exciton peaks of as-transferred \mos2 (light and dark blue curves) are slightly blue-shifted with respect to that of vacuum annealed \mos2 (orange curve) and lie at 1.887\rpm\SI{0.002}{\electronvolt} which agrees very well with the value of \SI{1.89}{\electronvolt} measured previously on \mos2 transferred from PDMS onto various substrates \cite{Buscema14}. On the contrary, for vacuum annealed \mos2 the A-exciton peak lies at 1.876\rpm\SI{0.002}{\electronvolt}, very close to that of \mos2 exfoliated directly on \sio2 \cite{Schuller13, Amani15}. By comparing the PL spectra at various locations on the two flakes in Figs.\ 4a, b and performing multi-Lorentzian fitting to decouple the trion peak from the exciton peak, we estimated a blue-shift of 11\rpm\SI{3}{\milli\electronvolt} for the A-exciton peak of as-transferred \mos2. Moreover, one can also notice a blue-shift in the B-exciton peak as highlighted by the Lorentzian fits in Fig.\ 4e. The complete set of fits for the entire spectra can be found in supplementary Fig.\ S6. \\[-1.5ex]



In order to understand the origin of the A, B-exciton blue-shift upon transfer, we performed Raman spectroscopy to gain further insight from the \e12g mode which is sensitive to strain in \mos2. In exfoliated, unstrained 1L-\mos2 at room temperature, the \e12g peak should lie at \SI{385}{\per\centi\meter} \cite{Li12} whereas for our as-transferred \mos2 on hBN it lies at \til\SI{386}{\per\centi\meter} (blue curve in Fig.\ 4f). An \e12g peak up-shift signifies an increase in built-in compressive strain, thus implying that PDMS not only leaves residues behind, but can also compress the \mos2 during transfer. An up-shift of \til\SI{1}{\per\centi\meter} corresponds to an accumulation of {\til}0.22\% compressive strain in \mos2 after transfer \cite{Conley13}, similar to estimate made by Buscema \textit{et al.} \cite{Buscema14} This induced strain causes an increase in the direct bandgap at the \emph{K}-point which leads to blue-shifted A, B-exciton emission \cite{Nayak15, Dou16}. According to previously reported DFT calculations \cite{Liu14b} as well as experimental results \cite{Conley13}, a 0.22\% strain would induce an A-exciton shift of 10\rpm\SI{1.5}{\milli\electronvolt} which is in good agreement with the shift of 11\rpm\SI{3}{\milli\electronvolt} estimated from our PL measurements. For the sake of completeness, one can also compare the \a1g peaks in the two Raman spectra to characterize the effect of residues on doping. The strong electron-phonon coupling of the out-of-plane \a1g mode in \mos2 causes it to down-shift with increasing doping \cite{Chakraborty12}. In Fig.\ 4f, the two \a1g peaks lying at \til\SI{406.5}{\per\centi\meter} imply a low n-doping in both \mos2 flakes on hBN (compared to 403-\SI{404}{\per\centi\meter} for 1L-\mos2 on \sio2) \cite{Buscema14, Li12}.\\[-1.5ex]

The origin of strain can possibly be attributed to the inherent lack of stiffness of PDMS. As discussed previously by Gomez \textit{et al.} \cite{Gomez14}, PDMS being soft can get slightly deformed during transfer by the pressure exerted on PDMS upon coming in contact with the target substrate. It is quite likely that this deformation of PDMS induces a compressive strain in the flake being transferred as measured for \mos2 in our case. The induced strain eventually gets released upon vacuum annealing, down-shifting the \e12g peak to \SI{385}{\per\centi\meter} and at the same time red-shifting the A, B-exciton peaks to unstrained values close to that of directly exfoliated \mos2. The release of transfer induced compression in \mos2 can also be evidenced in the AFM scans in Figs.\ 3c, d before and after annealing where an increase in the total surface area upon annealing can be clearly noticed. \\[-1.5ex]

\section{Conclusions}
To summarize, we have demonstrated that PDMS residues as well as compressive strain can be present in 2D materials exfoliated onto and transferred from PDMS stamps. Using 1L-\mos2 as an example, we have observed evidence of surface contamination in both AFM maps and PL spectra. \uvo3 treatment of PDMS prior to exfoliation significantly reduces the amount of unwanted surface oligomers which results in a cleaner transfer of \mos2 in comparison with untreated PDMS. We showed that a 1L-\mos2 surface of a very high quality can be obtained by a combination of \uvo3 pre-cleaning followed by vacuum annealing after transfer. AFM topography of annealed \mos2 flakes on hBN displayed a homogeneously smooth surface with a substantial reduction in interfacial bubbles and wrinkles. PL spectroscopy of as-transferred \mos2 revealed blue-shifted A, B-exciton peaks due to an accumulation of compressive strain during the transfer. This induced compression could be released by a post-transfer vacuum anneal. \\[-1.5ex]

It is advantageous to exfoliate \mos2 on PDMS for obtaining large area ({\small \gt}\SI{1000}{\upmu\meter\squared}) monolayer \mos2 flakes with a high yield unlike direct exfoliation on \sio2 which results in relatively smaller flakes. The recipe we provided now makes it possible to integrate these large area flakes exfoliated on PDMS into clean heterostructures for high performance electronic and photonic devices. Our results are valuable for future experimental studies and practical applications utilizing clean 2D material heterostructures. \\[-1.5ex]

One must keep in mind that even though PDMS residues do not significantly influence the optical properties of \mos2, surface residues could still affect electrical transport by scattering charge carriers and thereby reduce carrier mobility. Moreover, trapped residues within TMDC heterostructures could also weaken interlayer coupling \cite{Tongay14} and adversely affect physical phenomena such as interlayer charge recombination or separation, interlayer electron-phonon coupling, polariton formation, out-of-plane tunneling, etc. which rely on high-quality interfaces. Hence, the importance of eliminating residues while stacking 2D materials using PDMS, or any other technique in general, cannot be overstated. In this direction, our results highlight the significance of a careful characterization and optimization of any transfer procedure and subsequent processing steps in order to preserve interface quality and obtain unperturbed crystal structures with well-defined physical properties. \\[-1.5ex]

\section*{Methods}
All transfers were done with a S\"USS MicroTec MJB4 mask aligner. Annealing was performed in a quartz tube furnace (Carbolite Gero) at \SI{200}{\degreeCelsius} for \SI{3}{\hour} in a low vacuum of \SI{5e-3}{\milli\bar} (limited by our rotary pump). Before heating, the quartz tube was flushed several times with argon gas to remove any residual water or oxygen molecules. After \SI{3}{\hour}, the furnace was left for a few hours to cool down naturally to room temperature before taking out the samples. \\[-1.5ex]

All measurements were carried out at room temperature under ambient conditions. For PL measurements, samples were mounted on a piezoelectric stage and excited with a \SI{532}{nm} Nd:YAG laser (attenuated to \SI{15}{\upmu\watt} power) using a 50x air objective (0.8 NA) in a scanning confocal microscopy setup. The \mos2 flakes were raster scanned across the laser focus and photon counts at each xy-position were recorded with a single photon counting module (Perkin Elmer SPCM-AQRH-14) after passing through a \SI{532}{nm} RazorEdge longpass filter. PL spectra were measured with a Princeton Instruments Acton SP300i spectrometer at \SI{100}{\upmu\watt} excitation power (\SI{30}{\second} integration). For Raman spectroscopy, excitation was done with the \SI{530.8}{nm} line of an Ar-Kr laser (Coherent) at \SI{100}{\upmu\watt} power and the back-scattered light was dispersed onto a grating with 1200 grooves/mm. The grating was calibrated with a Neon lamp before recording the Raman spectra. Lorentzian fits were obtained using the freely available \textit{peak-o-mat} software. All spectra were normalized to the Raman peak of \textit{p$^+$}Si lying at \SI{521.7}{\per\centi\meter} for a better comparison. \\[-1.5ex]

\section*{Author Contributions}
The overall project was conceived and supervised by PB and LN. AJ developed the cleaning procedure, fabricated the samples, performed the measurements and wrote the manuscript. PB built the confocal optical microscopy setup used for PL spectroscopy of \mos2. Raman characterization was carried out together with SH. The vacuum tube furnace was setup by MP who also wrote the LabVIEW script for performing vacuum annealing. TT and KW synthesized the hBN crystals used in this study. PB, SH, MP and LN also participated in discussion of results and contributed to writing the manuscript. \\[-1.5ex]

\section*{Acknowledgments}
This research was supported by the Swiss National Science Foundation (grant no.\ 200021\_165841) and the ETH Z\"urich (ETH-32 15-1). TT and KW acknowledge support from the Elemental Strategy Initiative conducted by the MEXT, Japan and JSPS KAKENHI (grant no.\ JP15K21722). \\[-1.5ex]

\newpage



\section*{Supplementary material}

*Supplementary Information*

**Minimizing Residues and Strain in 2D Materials Transferred from PDMS**

Achint Jain,[1] Palash Bharadwaj,[2] Sebastian Heeg,[1] Markus Parzefall,[1]
Takashi Taniguchi,[3] Kenji Watanabe,[3] and Lukas Novotny[1]

[1]*Photonics Laboratory, ETH Zürich, 8093 Zürich, Switzerland*
[2]*Department of Electrical and Computer Engineering,
Rice University, Houston, TX 77005, USA*
[3]*National Institute for Material Science, 1-1 Namiki, Tsukuba 305-0044, Japan*

## List of Figures

## S1. Ineffectiveness of annealing in removing PDMS residues

Residues are always present to a varying degree on flakes transferred from PDMS. Here we show an exceptional case of residues on a 1L-MoS$_2$ flake which was exfoliated on untreated PDMS (Fig. S1a inset) and transferred to hBN (Fig. S1a). The AFM topography map of the as transferred flake in Fig. S1b clearly reveals that it is covered in a thick coat of PDMS residues in most places. As indicated by the red cross-section in Fig. S1c, the thickness of this residue layer is >5 nm near the MoS$_2$ edge. It is a common practice to anneal 2D materials to clean surface residues after transfer and enhance interlayer coupling in heterostructures. However, as shown in the main text, PDMS residues are difficult to completely remove simply by annealing. Figure S2d is an AFM map of the same flake after 3 h annealing at 200 °C in vacuum. Bubbles and wrinkles were eliminated as expected and the total thickness came down to ~1.6 nm (green profile in Fig. S2c) upon annealing. But this is still not comparable to the thickness of ~0.7 nm exhibited by monolayer MoS$_2$ transferred from UV-O$_3$ treated PDMS after annealing. These results demonstrate that substantial residues are frequently present on MoS$_2$ flakes transferred from untreated PDMS and can't be removed solely by annealing which highlights the need for a PDMS pre-cleaning technique in order to build cleaner heterostructures.

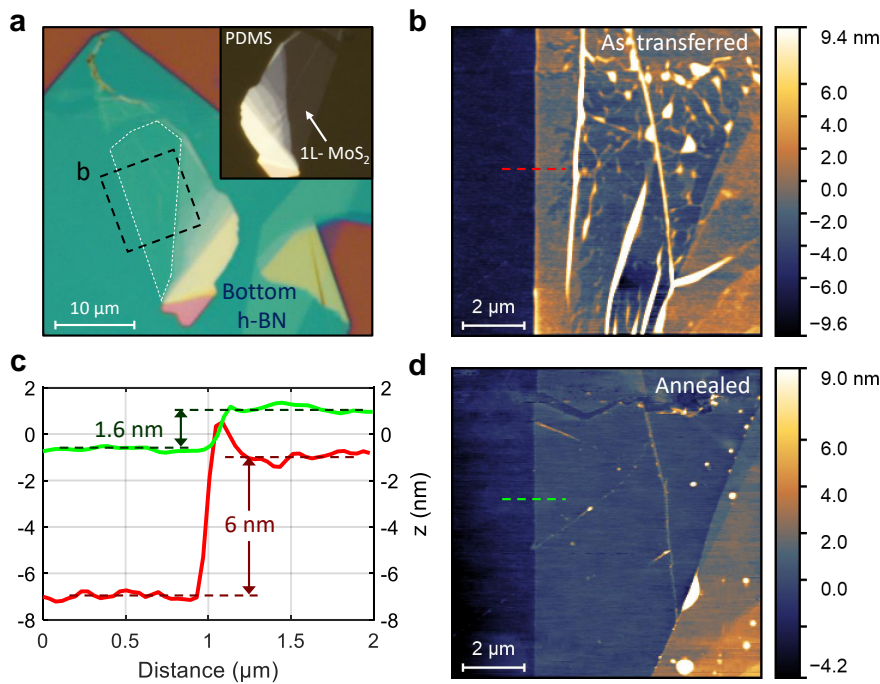

**Figure S1**. **Another MoS$_2$ flake with PDMS residues.** Optical image of another MoS$_2$ flake transferred onto hBN from untreated PDMS. The monolayer segment has been outlined. Inset: The same MoS$_2$ flake on PDMS before transfer. **(b)** AFM topography map of the region outlined in **a** displaying thick, inhomogeneous layers of residues covering most of the MoS$_2$ flake. **(c)** Height profile along the red and green dashed lines in **b, d** indicating the total thickness before and after annealing respectively. **(d)** AFM map of the same region after 3 h vacuum annealing. PDMS residues are still present as evident from the green profile although the flake is nearly free from bubbles/wrinkles.

Besides UV-O$_3$ treatment, we also tried other alternate approaches to clean the PDMS surface before exfoliation. One possible way to clean PDMS is via extraction of uncrosslinked oligomers from bulk PDMS by soaking it in organic solvents for several hours[1]. But we found that it is difficult to completely extract all oligomers this way[2] and being a lengthy procedure makes it inconvenient for practical use. Another alternative to UV-O$_3$ for breaking down organic species is O$_2$ plasma cleaning. However, we noticed that O$_2$ plasma can be too harsh for PDMS and could lead to the formation of fine cracks on the PDMS surface after few minutes exposure at 100 W power, similar to observations made by Bodas *et al.*[3]. Moreover, O$_2$ plasma treated PDMS surface gave us a very poor yield of flakes upon exfoliation as very few of them stuck to it compared to UV-O$_3$ treated PDMS. The O$_2$ plasma treatment process could possibly be optimized but we did not investigate it further. It is also important to mention that the choice of PDMS didn't make a big difference. We also tried PDMS films synthesized using Sylgard 184 (Dow Corning) but found similar residues as with Gel-Pak films.

## S2. Visualizing PDMS residues in an optical microscope

Under certain circumstances, PDMS residues on a Si/SiO$_2$ (285 nm) substrate can be even identified simply in an optical microscope. The apparent color of the Si/SiO$_2$ substrate as seen in reflection arises from interference and is quite sensitive to the thickness and refractive index of an additional dielectric layer on top. In case of MoS$_2$ transferred from PDMS to SiO$_2$ without the application of heat during the transfer (i.e. at room temperature), the residual PDMS layer can sometimes be thick enough to cause a detectable change in the color of the Si/SiO$_2$ substrate. Figure S3a is an optical microscope image of MoS$_2$ flakes exfoliated on untreated PDMS. The PDMS was brought in contact with SiO$_2$ and then slowly detached, without any intermediate heating step while in contact. Among the flakes visible in Fig. S3a, the large thick MoS$_2$ flake in the lower part did not get transferred to SiO$_2$. Interestingly, a clear outline of the missing flake on the SiO$_2$ substrate can be easily seen in Fig. S3b after transfer. In this region where the thick MoS$_2$ flake prevented the PDMS from coming in contact with the SiO$_2$, the substrate retained its original color whereas in areas where PDMS came in direct contact with SiO$_2$, transferred residues led to a change in color. This gave rise to a visible contrast between clean and PDMS contaminated SiO$_2$ surface which is easily noticeable in an optical microscope. For transfers done at 65 °C, PDMS residues are harder to see directly but they can still be visualized by differential interference contrast (DIC) microscopy (images not included here).

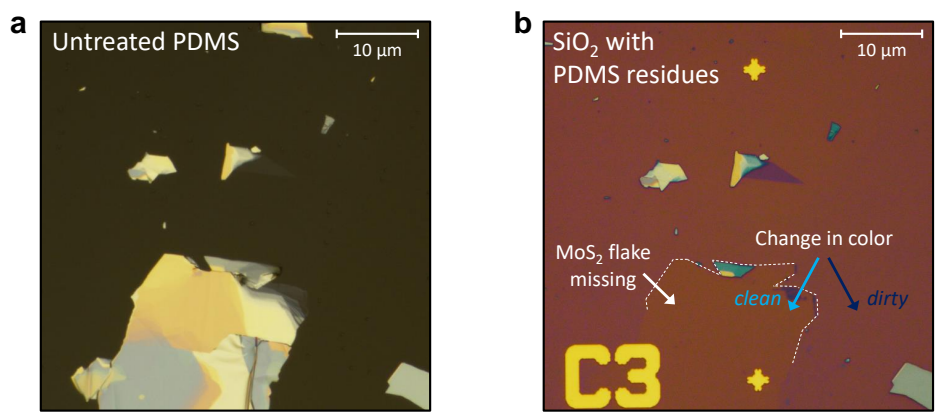

**Figure S2**. **Visualizing PDMS residues in an optical microscope.** (a) Optical image of MoS$_2$ flakes exfoliated on untreated PDMS. The image has been mirrored for an easier comparison with the right-hand side image. (b) Optical image of the same flakes after transfer to SiO$_2$. The large MoS$_2$ flake did not get transferred but its outline is still clearly visible. This is due to a change in color of the surrounding areas which came in direct contact with PDMS and got contaminated. The clean SiO$_2$ area has been partly demarcated to serve as a guide to the eye.

## S3. Additional data on clean transfer from UV-O$_3$ treated PDMS

In this section we show two more examples of clean MoS$_2$ flakes transferred using our new recipe. Figure S3a is an optical image of a 1L-MoS$_2$ flake exfoliated on UV-O$_3$ treated PDMS (inset) and transferred to hBN. The AFM topography map of the as-transferred flake in Fig. S3b shows a mostly clean surface (though slight traces of residues can be identified in the lower right part). Remarkably, even a monolayer hBN terrace can be resolved in this image which highlights the superior cleanliness of MoS$_2$ transferred this way compared to that from untreated PDMS in Fig. 1 of the main text. The higher resolution AFM map in Fig. S3c reveals a pristine surface (except for bubbles) with a clean step of 6.9 Å (Fig. S3d) and the corresponding AFM phase map in Fig. S3e displays strong phase contrast between MoS$_2$ and hBN. These images are quite unlike Figs. 1e-i where the step height and phase contrast were both obscured by residues. It is important to mention that these maps were recorded on as-transferred MoS$_2$ before annealing but are still remarkably clean. Figure S3f is an AFM topography map of the same flake after vacuum annealing showing a relaxed surface without even a slight trace of residues. Moreover, the MoS$_2$ in the lower

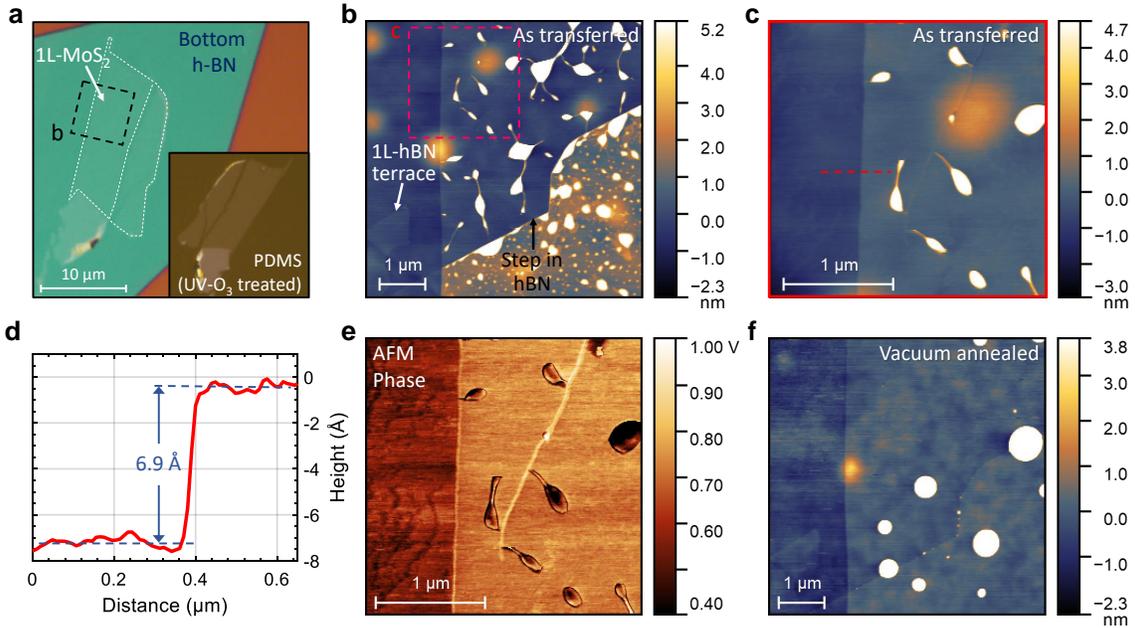

**Figure S3**. **An example of clean MoS$_2$ before annealing.** (**a**) Optical microscope image of another 1L-MoS$_2$ flake transferred from UV-O$_3$ treated PDMS. (**b**) AFM topography map of the as-transferred flake taken from the region outlined in **a**. Even a 1L-hBN terrace is resolvable in this image which is a clear evidence of a significantly cleaner surface compared to Fig. 1 of the main text. (**c**) Higher resolution AFM map of the region outlined in **b** showing the residue-free surface of the as-transferred MoS$_2$ without any annealing. (**d**) Height profile along the red dashed line in **c** displaying a clean step of 6.9 Å corresponding to pristine monolayer MoS$_2$. (**e**) AFM phase map corresponding to the topography in **c** exhibiting a strong phase contrast between MoS$_2$ and hBN which again points towards the absence of residues. (**f**) AFM map of the same region as in **b** after vacuum annealing.

right part makes much better contact with the hBN below after spreading out during annealing leading to an apparent reduction in thickness. Hence, this leads us to conclude that a combination of UV-O$_3$ pre-treatment followed by vacuum annealing is the optimum for obtaining a pristine MoS$_2$ surface with an unperturbed lattice.

In Fig. 2 of the main text, we only presented AFM results obtained from the left (triangular) 1L-MoS$_2$ flake in Fig. 2b after transfer from UV-O$_3$ treated PDMS. Here we have included additional data from the right (trapezoidal) 1L-MoS$_2$ attached to the same flake. AFM maps before and after annealing are depicted in Fig. S4a, b respectively and exhibit a smooth, homogeneous, residue-free surface after annealing in contrast to Figs. 2a and S2d. In the high-resolution map in Fig. S4c, a pristine MoS$_2$ surface with a thickness of 8.5 Å (Fig. S4d) can be clearly seen. These results in line with those in Fig. 3 and highlight the effectiveness of our UV-O$_3$ pre-cleaning process in reducing residues.

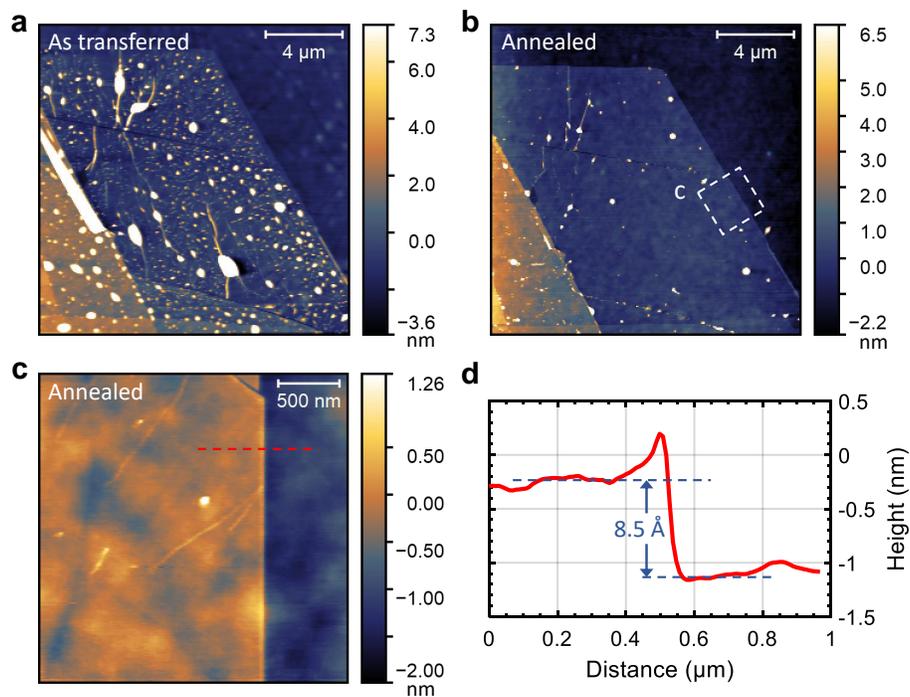

**Figure S4. Additional example of clean MoS$_2$ transfer.** (**a**) AFM map of the right 1L-MoS$_2$ flake in Fig. 2b of the main text. (**b**) AFM map of the same area after vacuum annealing. (**c**) Higher resolution map of the region outlined in **b** exhibiting a pristine surface. The small undulations in topography are not variations in the MoS$_2$ thickness but arise from the substrate below. (**d**) Height profile along the red dashed line in **c** displaying a thickness (8.5 Å) very close to that of monolayer MoS$_2$.

## S4. Effect of annealing on MoS$_2$ lying on SiO$_2$

So far we have presented several examples demonstrating efficient removal of bubbles from the MoS$_2$-hBN interface via vacuum annealing. However, annealing doesn't seem to be very effective in getting rid of bubbles accumulated at the MoS$_2$-SiO$_2$ interface. Figure S5a is an optical image of a long MoS$_2$ flake transferred from UV-O$_3$ treated PDMS onto a Si-SiO$_2$ substrate (pre-cleaned with O$_2$ plasma). AFM map of the as transferred 1L-MoS$_2$ region in Fig. S5b reveals a high density of tiny bubbles trapped between MoS$_2$ and SiO$_2$. Unfortunately, after 3 h vacuum annealing at 200 °C, the distribution of bubbles in Fig. S5c looks quite similar to that before annealing, unlike in case of MoS$_2$ on hBN. To understand these contrasting behaviors, we must take into account the strong adhesion between MoS$_2$ and SiO$_2$ due to Coulombic attraction arising from dangling bonds and surface charges on SiO$_2$. Moreover, it can be noticed in the high-resolution map in Fig. S5d that the bubbles on SiO$_2$ are much smaller in size than those seen previously in MoS$_2$ on hBN (cf. Fig. 1e) which also points towards a higher adhesion energy for MoS$_2$ on SiO$_2$[4,5]. This strong adhesion together with the increased sliding friction[6] caused by the higher surface roughness of SiO$_2$ keeps MoS$_2$ anchored in place during annealing. On the other hand, only a weak van der Waals attraction exists between MoS$_2$ and hBN which allows a greater freedom of movement for

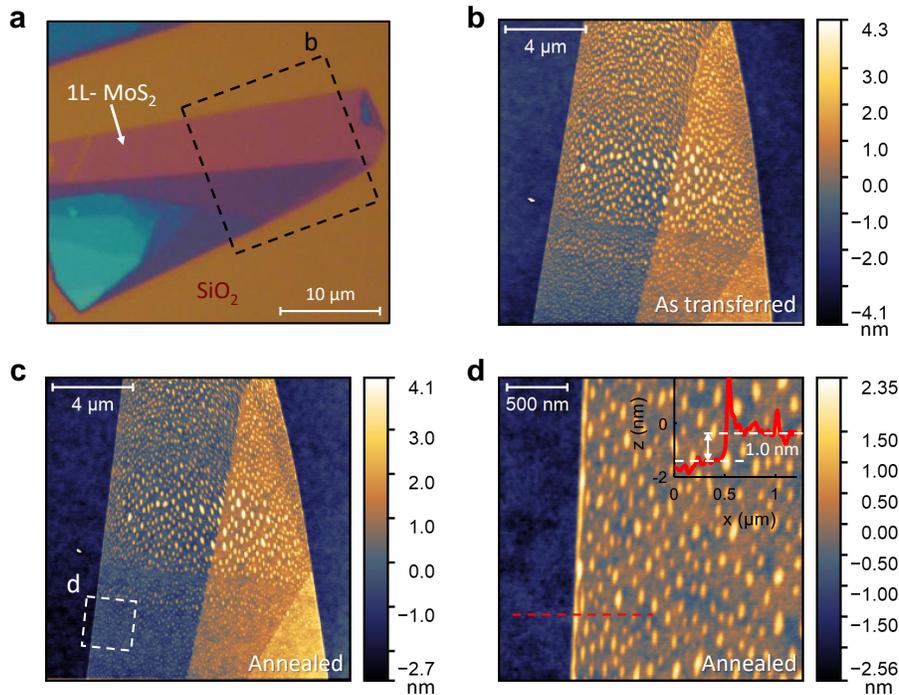

**Figure S5. Bubbles at the MoS$_2$-SiO$_2$ interface.** (a) Optical microscope image of a 1L-MoS$_2$ flake transferred to SiO$_2$ (285 nm) from UV-O$_3$ treated PDMS. (b) AFM topography map of the black outlined region in **a** displaying a high density of tiny bubbles. (c) AFM map of the same region after 3 h vacuum annealing with the distribution of bubbles largely unaffected by annealing. (d) Higher resolution scan of the white outlined region in **c**. Inset: height profile along the red dashed line

MoS$_2$. Moreover, the reduced sliding friction[6] on the atomically smooth hBN surface makes it easier for ripples in the MoS$_2$ layer and mobile species trapped inside bubbles to slide out during annealing which would otherwise have been immobilized on SiO$_2$. This behavior illustrates that the choice of substrate plays a crucial role in governing the surface morphology of transferred flakes and deserves greater attention in future studies.

## S5. Fitting the photoluminescence spectra

Here we show the complete set of fits for the photoluminescence (PL) spectra plotted in Fig. 3d of the main text. In Figs. S6a and b, the dark blue and orange data points were fitted with a model comprising of a sum of 4 and 3 Lorentzian functions respectively, with each Lorentzian representing one spectral feature. From the fits, a shift of 10 meV in the A-exciton peak position was deduced for this pair of spectra. The trion peak showed a much smaller shift of ~3 meV.

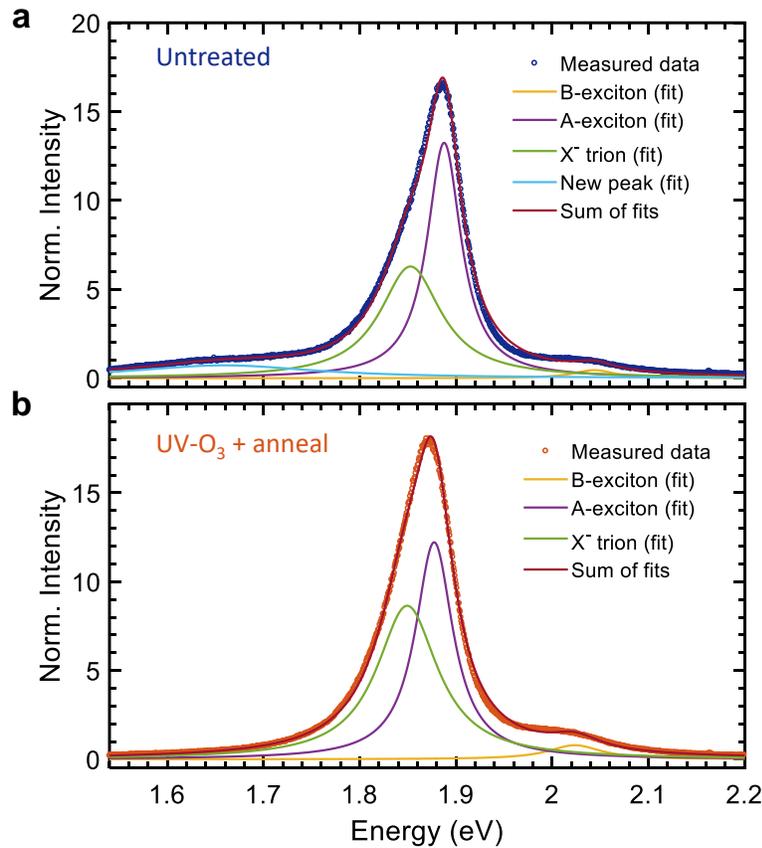

**Figure S6**. **Lorentzian fits to the PL spectra.** **(a)** Sum of four Lorentzians (red) fitted to the PL spectra of MoS$_2$ transferred from untreated PDMS. Along with the fitted sum, the individual fit components are also plotted separately. **(b)** Sum of three Lorentzians (red) fitted to the PL spectra of MoS$_2$ transferred from UV-O$_3$ treated PDMS and annealed. The empty circles are experimentally measured data points (reproduced from Fig. 3d of the main text) while the smooth curves are fits to the measured data.



\end{document}